\documentclass[twocolumn]{aastex62}

\defcitealias{finley2018dipquadoct}{FM18}
\usepackage{mhchem}

\usepackage{array,graphicx}
\usepackage{booktabs}
\usepackage{pifont}
\newcommand*\rot{\rotatebox{90}}

\graphicspath{{./}{figures/}}

\received{--}
\revised{--}
\accepted{--}

\shorttitle{Millennia of Solar Wind Torque}
\shortauthors{Finley et al.}

\begin{document}

\title{Solar Angular Momentum Loss Over the Past Several Millennia}

\correspondingauthor{Adam J. Finley}
\email{*af472@exeter.ac.uk}

\author{Adam J. Finley}
\affil{University of Exeter,
              Exeter, Devon, EX4 4QL, UK}
\author{Siddhant Deshmukh}
\affil{University of Exeter,
              Exeter, Devon, EX4 4QL, UK}
\author{Sean P. Matt}
\affil{University of Exeter,
              Exeter, Devon, EX4 4QL, UK}
\author{Mathew Owens}
\affil{University of Reading,
              Reading, Berkshire, RG6 6BB, UK}
\author{Chi-Ju Wu}
              \affil{Max-Planck-Institut f\"{u}r Sonnensystemforschung,
              Justus-von-Liebig-Weg 3, G\"{o}ttingen, Germany}



\begin{abstract}
The Sun and Sun-like stars lose angular momentum to their magnetised stellar winds. This braking torque is coupled to the stellar magnetic field, such that changes in the strength and/or geometry of the field modifies the efficiency of this process. Since the space-age, we have been able to directly measure solar wind properties using in-situ spacecraft. Furthermore, indirect proxies such as sunspot number, geomagnetic indices, and cosmogenic radionuclides, constrain the variation of solar wind properties on centennial, and millennial timescales.  We use near-Earth measurements of the solar wind plasma and magnetic field to calculate the torque on the Sun throughout the space-age. Then, reconstructions of the solar open magnetic flux are used to estimate the time-varying braking torque during the last nine millennia. We assume a relationship for the solar mass loss rate based on observations during the space-age which, due to the weak dependence of the torque on mass loss rate, does not strongly affect our predicted torque. The average torque during the last nine millennia is found to be $2.2\times10^{30}$erg, which is comparable to the average value from the last two decades. Our dataset includes grand minima (such as the Maunder Minimum), and maxima in solar activity, where the torque varies from $\sim1-5\times10^{30}$erg (averaged on decadal timescales), respectively. We find no evidence for any secular variation of the torque on timescales of less than $9000$ years.

\end{abstract}



\section{Introduction}

The observed rotation periods of most low-mass stars ($M_*\lesssim 1.3M_{\odot}$) on the main sequence can be explained by their magnetised stellar winds. These winds efficiently remove angular momentum causing stars to spin-down with age \citep{skumanich1972time, soderblom1983rotational, barnes2003rotational, barnes2010simple, delorme2011stellar, van2013fast, bouvier2014angular}. Throughout this process, their magnetic field generation (due to the dynamo mechanism) is strongly linked with rotation \citep{brun2017magnetism}, and the strength of the magnetic field is found to influence the efficiency of angular momentum transfer through the stellar wind \citep{weber1967angular, mestel1968magnetic, kawaler1988angular, matt2012magnetic, garraffo2015dependence}. The resulting strong dependence of torque on rotation rate leads to a convergence of rotation periods with age, as initially fast rotating stars generate strong magnetic fields and experience a larger braking torque than the initially slowly rotating stars. This spin-down is also observed to be a function of stellar mass \citep{agueros2011factory, mcquillan2013measuring, nunez2015linking, rebull2016rotation, covey2016rapidly, agueros2017setting, douglas2017poking}.

Many models now exist to study the rotation period evolution of low-mass stars \citep{gallet2013improved, brown2014metastable, gallet2015improved, matt2015mass, johnstone2015stellar, amard2016rotating, blackman2016minimalist, sadeghi2017semi, see2018open, garraffo2018revolution}. Such models provide insight on how stellar wind torques evolve on secular timescales ($\sim$ Gyrs), independently from our understanding of the braking mechanism. For Sun-like stars, the torques prescribed by these models are averaged over fractions of the braking timescale ($\sim10-100$Myrs). However, we observe variability in the magnetic field of the Sun on a range of much shorter timescales \citep{derosa2012solar, vidotto2018magnetic}, which is expected to influence the angular momentum loss rate in the solar wind \citep{pinto2011coupling, reville2017global, finley2018effect, perri2018simulations}.

In \cite{finley2018effect}, the short timescale variability (from $\sim27$ days up to a few decades) of the solar wind was examined using in-situ observations of the solar wind plasma and magnetic field. By applying a braking law derived from MHD simulations by \cite{finley2018dipquadoct}, they calculated the time-varying torque on the Sun due to the solar wind. When averaged over the last $\sim 20$ years they found a solar wind torque of $2.3\times 10^{30}$erg. This value is in agreement with previous in-situ and data driven calculations \citep{pizzo1983determination, li1999magnetic}, and also recent simulation results \citep{alvarado2016simulating,reville2017global,fionnagain2018solar,usmanov2018steady}.

When compared to the torques required by rotation-evolution models \citep[e.g.][]{matt2015mass}, current estimates of the solar wind torque are smaller by a factor of $\sim 3$ {(this discrepancy was noted already by \citealp{soderblom1983rotational})}. {One possible explanation for the discrepancy is that the solar wind torque is variable, and that the torque is currently in a ``low state,'' or that the torque has recently, but permanently weakened (e.g., as suggested by \citealp{van2016weakened, garraffo2018revolution, o2018solar}).  For this to be true, the variations in the torque must have happened on timescales much longer than the space age (decades), but shorter than the timescales on which the rotation-evolution models are sensitive ($\sim 10^{8}$ years, for solar-aged stars).}

{In this work, we employ reconstructions of solar wind properties from the literature, in order to estimate the solar wind torque further back in time than has been probed so far (more than two orders of magnitude).  Although we still cannot probe the timescales of rotational evolution, this helps to elucidate the types of variability that may occur in the solar wind torque.}
 We first describe the \cite{finley2018dipquadoct} braking law, hereafter \citetalias{finley2018dipquadoct}, in Section 2. Then we estimate the angular momentum loss rate, due to the solar wind, through the space age using in-situ data in Section 3. Finally, in Section 4, we use reconstructions of the Sun's open magnetic flux (which are based on sunspot number, geomagnetic indices, and cosmogenic radionuclide records), to estimate the angular momentum loss rate on centennial and millennial timescales.


\section{Angular Momentum Loss Formulation}


Generally, the torque on a star due its magnetised wind {can be written}
\begin{equation}
    \tau=\dot{M}\Omega_*R_*^2\bigg(\frac{\langle R_A\rangle}{R_*}\bigg)^2,
    \label{tau}
\end{equation}
where $\dot{M}$ is the mass loss rate, $\Omega_*$ is the stellar rotation rate, $R_*$ is the stellar radius, and $\langle R_A\rangle/R_*$ can be thought of as an efficiency factor for the angular momentum loss rate which, under the assumption of ideal steady-state MHD, scales as the average Alfv\'en radius \citep{weber1967angular, mestel1968magnetic}.

We use a semi-analytic formula for $\langle R_A\rangle$ which depends on the open magnetic flux, $\phi_{open}$, and mass loss rate, $\dot M$, in the wind (\citealp{reville2015effect}; \citealp{strugarek2014modelling};  \citealp{reville2015solar}; \citealp{reville2016age}; \citealp{pantolmos2017magnetic}; \citealp{finley2017dipquad}; \citetalias{finley2018dipquadoct}). We define the open magnetic flux as the total unsigned flux that permeates the stellar wind,
\begin{equation}
    \phi_{open}=\oint_A|{\bf B}\cdot d{\bf A}|,
\end{equation}
where $\bf B$ is the magnetic field strength in the wind, and A is a closed surface which is located outside the last closed magnetic field line. In a steady-state, the last closed magnetic field line resides within the Alfv\'en radius, $R_A$, which is defined as the location where the wind speed becomes equal to the Alfv\'en speed, $v(R_A)=v_A=B_A/\sqrt{4\pi\rho_A}$, where the subscript $A$ denotes values taken at $R_A$. Considering a steady, MHD flow, along a one dimensional magnetic flux tube, mass and magnetic flux are conserved. Therefore, in a steady-state stellar wind, where the flow is spherically symmetric, the magnetic field strength at $R_A$ is specified by flux conservation as $B_A=\phi_{open}/(4\pi R_A^2)$. The Alfv\'en speed is then,
\begin{equation}
v_A^2=\frac{\phi_{open}^2/(4\pi)^2 R_A^4}{4\pi\rho_A},
\end{equation}
which by rearranging, and then substituting for $\dot M$, produces a relation for $R_A$,
\begin{equation}
\label{eq_raanalytic}
R_A^2=\frac{\phi_{open}^2}{(4\pi)^2v_A (4\pi\rho_A v_A R_A^2) } = \frac{\phi_{open}^2}{(4\pi)^2 v_A \dot M }.
\end{equation}
{Since real stellar winds are multi-dimensional in nature, several authors (e.g., \citealp{matt2008accretion}; \citealp{pinto2011coupling}; \citealp{matt2012magnetic}; \citealp{cohen2014grid}; \citealp{reville2015effect}; \citealp{reville2015solar}; \citealp{garraffo2016missing}; \citealp{pantolmos2017magnetic}; \citealp{finley2017dipquad}; \citetalias{finley2018dipquadoct}) have employed magnetohydrodynamic (MHD) numerical simulations to derive semi-analytic scalings for the wind torques.  A few of these studies have derived a relationship similar to equation (\ref{eq_raanalytic}), which has the form}
\begin{equation}
\frac{\langle R_A\rangle}{R_*}=K\bigg[\frac{\phi_{open}^2/R_*^2}{\dot M v_{esc}}\bigg]^m,
\label{OG_eq}
\end{equation}
where $\langle R_A\rangle/R_*$ is calculated from the simulations by inverting equation (\ref{tau}), and $K$ and $m$ are fit constants. In equation (\ref{OG_eq}){, compared to equation (\ref{eq_raanalytic})}, $v_A$ has been replaced by the surface escape speed, $v_{esc}=\sqrt{2GM_{\odot}/R_{\odot}}$, and any dependence $v_A$ has on $\phi_{open}$ and $\dot M$ is absorbed into the fit constants. These fit constants also account for the multiplicative factor of $(4\pi)^2$, and any effects introduced by the flow being multi-dimensional in nature.
{The formulation of equation (\ref{OG_eq})} for $\langle R_A\rangle$, using $\phi_{open}$, {is} insensitive to how the coronal magnetic field is {structured} {(i.e., insensitive to the geometry of the magnetic field; \citealp{reville2015effect}), but the fit constants can} be affected by differing wind acceleration profiles \citep{pantolmos2017magnetic}, and 3D structure in the mass flux.

{We adopt the fit parameters from \citetalias{finley2018dipquadoct}.} For the Sun, equation (\ref{OG_eq}) {then} reduces to,
\begin{eqnarray}
    \langle R_A\rangle=(12.9R_{\odot})\bigg(\frac{\dot M}{1.1\times 10^{12} [\text{g/s}]}\bigg)^{-0.37} \nonumber \\
    \times \bigg(\frac{\phi_{open}}{8.0\times 10^{22}[\text{Mx}]}\bigg)^{0.74},
    \label{open_ra}
\end{eqnarray}
using values of the solar mass, $M_{\odot}=1.99\times10^{33}$g, and radius, $R_{\odot}=6.96\times10^{10}$cm. For the solar wind torque, equation (\ref{tau}) becomes,
\begin{eqnarray}
    \tau =(2.3\times10^{30}[\text{erg}])\bigg(\frac{\dot M}{1.1\times 10^{12} [\text{g/s}]}\bigg)^{0.26} \nonumber \\
    \times \bigg(\frac{\phi_{open}}{8.0\times 10^{22}[\text{Mx}]}\bigg)^{1.48},
    \label{open_torque}
\end{eqnarray}
using the solar rotation rate $\Omega_{\odot}=2.6\times10^{-6}$rad s$^{-1}$. The torque depends only on $\phi_{open}$ and $\dot M$, given the choice of polytropic base wind temperature used in \citetalias{finley2018dipquadoct}. By comparing feasible base wind temperatures, \cite{pantolmos2017magnetic} showed there is at most a factor of $\sim 2$ difference in the prediction of equation (\ref{open_torque}) between the coldest and hottest polytropic winds (1.3-4.2MK for the Sun). The simulations of \citetalias{finley2018dipquadoct}, from which we derived equations (\ref{open_ra}) and (\ref{open_torque}), correspond to a base wind temperature of $\sim 1.7$MK, which sits at the lower edge of this temperature range (where the torques are strongest).

\section{Solar Wind Torque During the Space-Age}

\begin{figure}
 \centering
  \includegraphics[trim=1cm 0cm 0.6cm 0cm, width=0.45\textwidth]{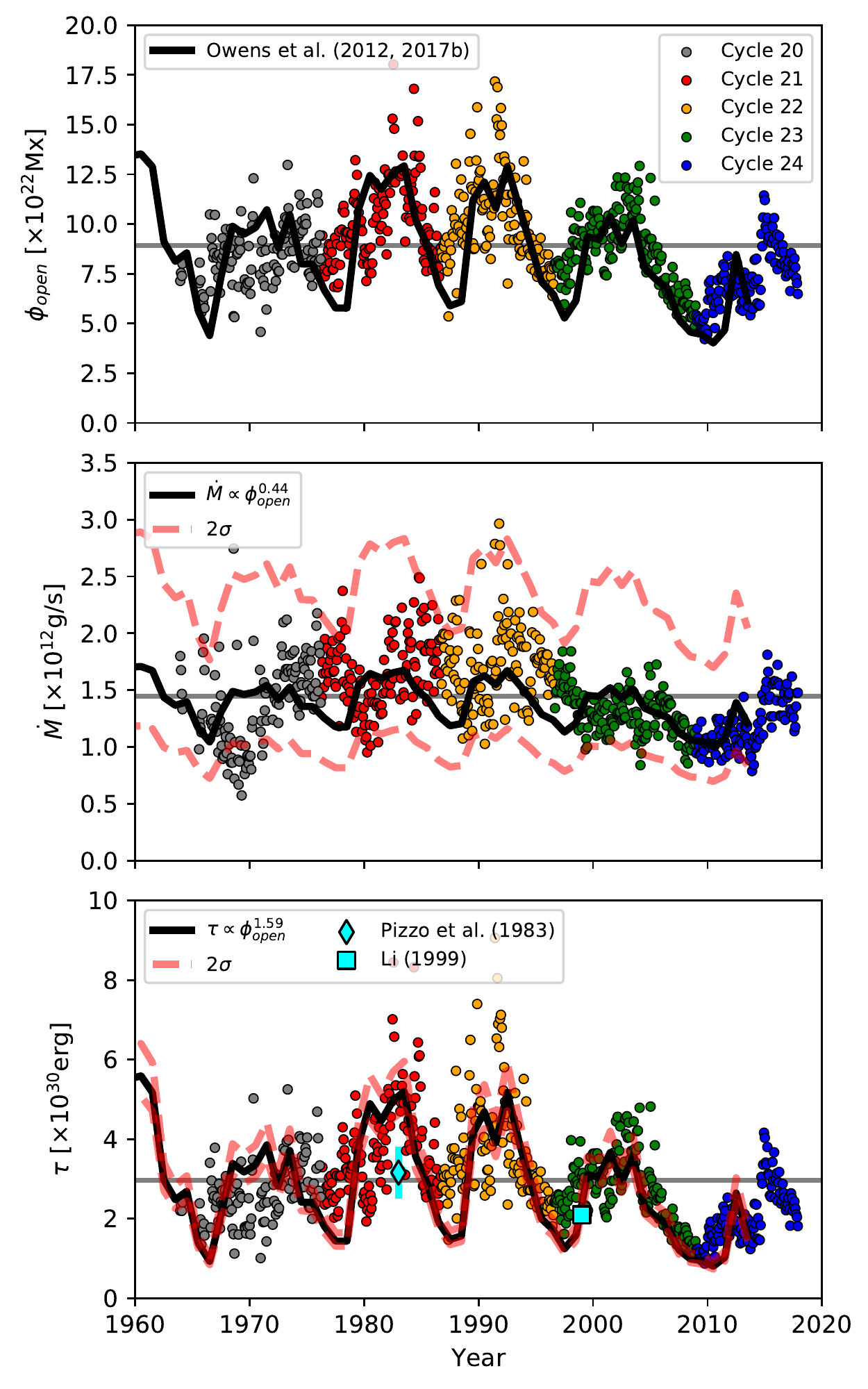}
   \caption{Several decades of open magnetic flux, $\phi_{open}$, and mass loss rate, $\dot{M}$, estimated from the OMNI dataset (near-Earth measurements), are shown with circles (color-coded by sunspot cycle number, 20-24) in the top two panels. The predicted solar wind torque, $\tau$, using equation (\ref{open_torque}) is then shown in the bottom panel. Averages of these three quantities are shown with grey horizontal lines. Over-plotted in each panel are the $\phi_{open}$ reconstruction from \cite{owens2017global}, the $\dot{M}$ predicted by equation ({\ref{MLR_eq}}), and the $\tau$ from equation (\ref{open_torque2}), with solid black lines. The 2$\sigma$ bounds for the predicted $\dot M$ and $\tau$, are indicated with dashed red lines.}
   \label{OMNI_data}
\end{figure}

\subsection{Observed Solar Wind Properties}
Hourly near-Earth solar wind plasma and magnetic field measurements are available from the OMNIWeb service\footnote{https://omniweb.gsfc.nasa.gov/ (Accessed in July 2018.)}. The OMNI dataset is compiled from the in-situ observations of several spacecrafts, from 1963 to present. We use measurements of the solar wind to estimate the open magnetic flux using,
\begin{equation}
    \phi_{open}=4\pi\langle R^2|B_R(R)|_{\text{1 hr}}\rangle_{\text{27 days}},
    \label{OSF_eq}
\end{equation}
where we average the radial magnetic field $B_R$, (taken from a single observing location) at a distance $R$ from the Sun, over a full solar rotation (27 days), and assume that the solar wind is roughly isotropic on our averaging timescale, in order to estimate the open magnetic flux. \cite{smith1995ulysses} were able to show that $R^2|B_R(R)|$ is approximately independent of heliographic latitude, as the solar wind is thought to redistribute significant variations in magnetic flux due to latitudinal magnetic pressure gradients caused by non-isotropy \citep{wang1995solar, lockwood2004open, pinto2017multiple}. Subsequently, the use of a single point measurement to infer the global open magnetic flux has been shown to be a reasonable approximation at distances less than $\sim 2$au by \cite{owens2008estimating}.

The open magnetic flux calculated using equation (\ref{OSF_eq}), during the space-age, is plotted in the top panel of Figure \ref{OMNI_data}. The 27 day averages are shown with circles that are colored according to the different sunspot cycles in our dataset. The average of this dataset is indicated with a grey horizontal line. The open magnetic flux roughly declines in time {over the past 3 cycles}, with the current sunspot cycle hosting some of the weakest values recorded in the OMNI dataset. Due to kinematic effects {that occur} between the Alfv\'en surface and the measurements taken at 1au, our estimate of the open magnetic flux is likely an upper limit \citep{owens2017sunward}.

Similarly to equation (\ref{OSF_eq}) for the open magnetic flux, the solar mass loss rate is estimated from in-situ measurements using,
\begin{equation}
    \dot M = 4\pi\langle R^2v_R(R)\rho(R)\rangle_{\text{27 days}},
    \label{MLR}
\end{equation}
{which is plotted in the middle panel of Figure~\ref{OMNI_data}.  Equation (\ref{MLR})} assumes the mass flux evaluated at a single observing location in the solar wind is representative of all latitudes when averaged over 27 days.
Using data from the fast latitude scans of the \textit{Ulysses} spacecraft, {\citet{finley2018effect} showed that the calculation of $\dot M$ from equation (\ref{MLR}) varies by a few 10's of percent when the spacecraft was emersed in slow, versus fast, solar wind streams}
 \citep[see also][]{phillips1995ulysses}. {Thus, the errors due to latitudinal variability are comparable to, but appear somewhat smaller than, the time-variability (see, e.g., \citealp{mccomas2013weakest}).  The cyclical variations of $\dot M$ are less clear than for the open flux, but they show}
 a similar decreasing trend {over the past 3 cycles.}

\subsection{Coronal Mass Ejections}
Equations (\ref{OSF_eq}) and (\ref{MLR}) do not take into account the effects of Coronal Mass Ejections (CMEs) in the data. These appear as impulsive changes (generally increases) in the observed solar wind properties, and clearly violate the assumed isotropy of wind conditions in equations (\ref{OSF_eq}) and (\ref{MLR}). CMEs occur once every few days at solar minimum, however their occurrence rate tracks solar activity, and at solar maximum they are observed on average five times a day \citep{webb2017there, mishra2019mass}. Previous authors have removed these events through the use of CME catalogues \citep{cane2003interplanetary} or clipping anomalous spikes \citep{cohen2011independency}. CMEs carry only a few percent of the total solar mass loss rate \citep{cranmer2017origins}, however at solar maximum they can provide a significant fraction of the average mass flux in the equatorial solar wind \citep{webb1994solar}.

\cite{finley2018effect} examined the effect of removing periods of high wind density ($>10$ cm$^3$) and high magnetic field strength ($>10$ nT), thought to correspond to the CMEs. They determined that the average open magnetic flux and mass loss rate, over their $\sim20$ years of data, decreased by $\sim4\%$ after these cuts were applied. As the role of CMEs in removing angular momentum is still in question \citep[see, e.g.][]{aarnio2012mass}, and their inclusion here is limited to a few percent, we present our results using the full unclipped dataset.

\subsection{Decades of Solar Wind Torque}
We use the open magnetic flux and mass loss rate estimates from Section 3.1 to compute the angular momentum loss rate in the solar wind using equation (\ref{open_torque}). The results from this calculation are shown in the bottom panel of Figure \ref{OMNI_data}. We calculate the average torque on the Sun during the space-age to be $2.97\times10^{30}$erg, which is larger than the value obtained by \cite{finley2018effect} of $2.3\times 10^{30}$erg due to the fact that \cite{finley2018effect} only examined the past $\sim20$ years. Averaging over each individual sunspot cycle, we find values of $2.67\times10^{30}$erg, $3.66\times10^{30}$erg, $3.70\times10^{30}$erg, $2.69\times10^{30}$erg, and $2.06\times10^{30}$erg, for cycles 20-24 respectively. Using equation (\ref{open_ra}), $\langle R_A \rangle$ is calculated to have its largest value in cycle 21 of $20.4R_{\odot}$, and minimum value of $7.7R_{\odot}$ in cycle 22. The value of $\langle R_A \rangle$ during the current sunspot cycle ranges from $\sim 8-16R_{\odot}$.


%

The time-varying torque computed here is in agreement with previous calculations of the solar wind torque. From the in-situ measurements of \cite{pizzo1983determination} using the \textit{Helios} spacecraft, to the recalculation of \cite{li1999magnetic} based on data from the \textit{Ulysses} spacecraft. Both of these estimates agree within the scatter of the 27 day averages computed in this work.

\section{Solar Wind Torque on Centennial and Millennial Timescales}
Up until now, we have examined only direct measurements of the solar wind. These observations have been facilitated by the exploration of near-Earth space, which began a few decades ago. For the centuries and millennia before this, only indirect measurements are available, such as sunspot observations \citep{clette2014revisiting}, measurements of geomagnetic activity \citep{echer2004long}, and studies of cosmogenic radionuclides found in tree rings or polar ice cores \citep[][]{usoskin2017history}. These indirect measurements are used to estimate longer time variability of the Sun's open magnetic flux \citep{lockwood2004open,vieira2010evolution,owens2011open,wusolar}. However these indirect measurements have limitations. Significantly for this work, they do not produce estimates for how the mass loss rate of the Sun has varied.

In this Section we produce a relation for the mass loss rate of the Sun, in terms of the open magnetic flux. Which is constructed using the range of observed values from Section 3.1. We then use this prescription for the mass loss rate, and equation (\ref{open_torque}), to evaluate the torque on the Sun due to the solar wind based on {indirect} reconstructions of the open magnetic flux.

\subsection{Estimating the Mass Loss Rate, and Wind Torque with the Open Magnetic Flux}
\begin{figure}
\centering
 \includegraphics[trim=0.7cm 0cm 0.2cm 0cm, width=0.45\textwidth]{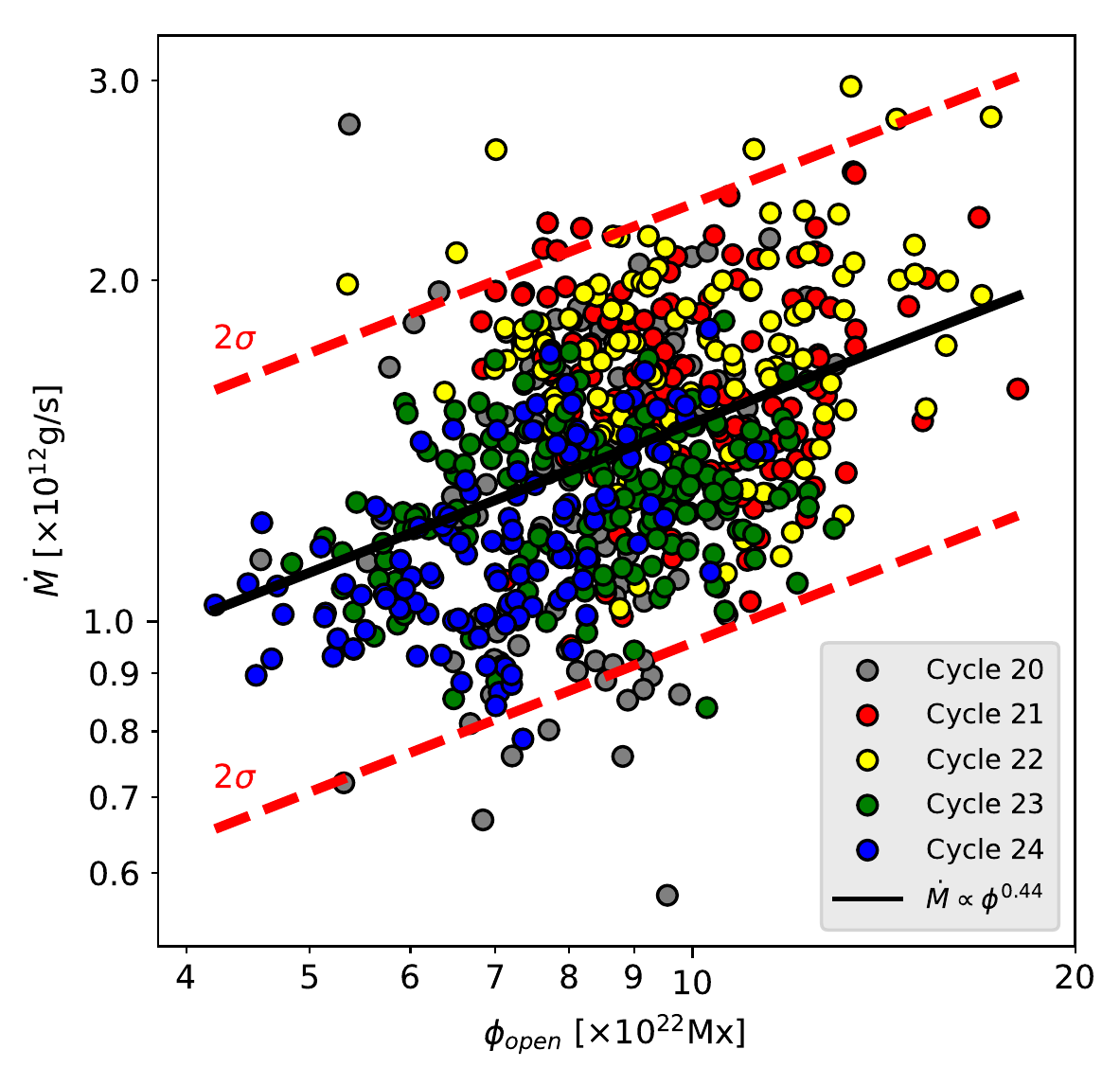}
  \caption{Mass loss rate, $\dot{M}$, versus open magnetic flux, $\phi_{open}$, derived the in-situ observations of the OMNI dataset. Values are color-coded by sunspot cycle, 21-24. The black line corresponds to the power law fit of equation (\ref{MLR_eq}). The dashed red lines indicates the 2$\sigma$ bounds given by a log-gaussian centered on the fit line.}
  \label{Mdot_OSF}
\end{figure}

Predicting the mass loss rates for low-mass stars, such as the Sun, is a difficult challenge, which has been attempted by previous authors to varying success \citep{reimers1975circumstellar, reimers1977absolute, mullan1978supersonic, schroder2005new, cranmer2011testing, cranmer2017origins}. The mass loss rates from Section 3.1 are plotted against their respective open magnetic flux values in the top panel of Figure \ref{Mdot_OSF}, colored by sunspot cycle. A weak trend of increasing mass loss rate with increasing open magnetic flux is observed. We fit a power-law relation for the mass loss rate in terms of the open magnetic flux,
\begin{equation}
  \dot M_{fit} =  (1.26\times 10^{12}[\text{g/s}])\bigg(\frac{\phi_{open}}{8.0\times 10^{22}[\text{Mx}]}\bigg)^{0.44},
  \label{MLR_eq}
\end{equation}
which is plotted as a solid black line.

{There is a large scatter around the fit of equation (\ref{MLR_eq}), which we wish to propagate through our calculation. We show the 2$\sigma$ limits of a log-gaussian function, centered on the fit, with red dashed lines. These lines are given by $\dot M^{-}_{fit}=0.64\dot M_{fit}$, and $\dot M^{+}_{fit}=1.57\dot M_{fit}$. When we estimate the mass loss rate for the historical estimates of the open magnetic flux in Sections 4.3, we will use both equation (\ref{MLR_eq}) and the 2$\sigma$ bounds.}

With the mass loss rate prescribed in terms of the open magnetic flux, we simplify equation (\ref{open_torque}) further to,
\begin{equation}
  \tau =  (2.4\times10^{30}[\text{erg}])\bigg(\frac{\phi_{open}}{8.0\times 10^{22}[\text{Mx}]}\bigg)^{1.59},
  \label{open_torque2}
\end{equation}
where the solar wind torque is now given solely as a function of open magnetic flux. Similarly, the 2$\sigma$ bound of equation (\ref{MLR_eq}) is propagated through equation (\ref{open_torque}) to give, $\tau^{-}=0.89\tau(\phi_{open})$, and $\tau^{+}=1.12\tau(\phi_{open})$. This allows us to predict the torque on the Sun due to the solar wind solely from the value of the open magnetic flux.  {Note that large ($\sim 50$\%) uncertainties in $\dot M$ translates to only a $\sim 10$\% uncertainty in torque, due to the weak dependence of $\tau$ on $\dot M$ in equation (\ref{open_torque}).}

\subsection{Reconstructions of the Solar Open Magnetic Flux}

\begin{figure*}
\centering
 \includegraphics[trim=2.5cm 1cm 2.5cm 0cm, width=\textwidth]{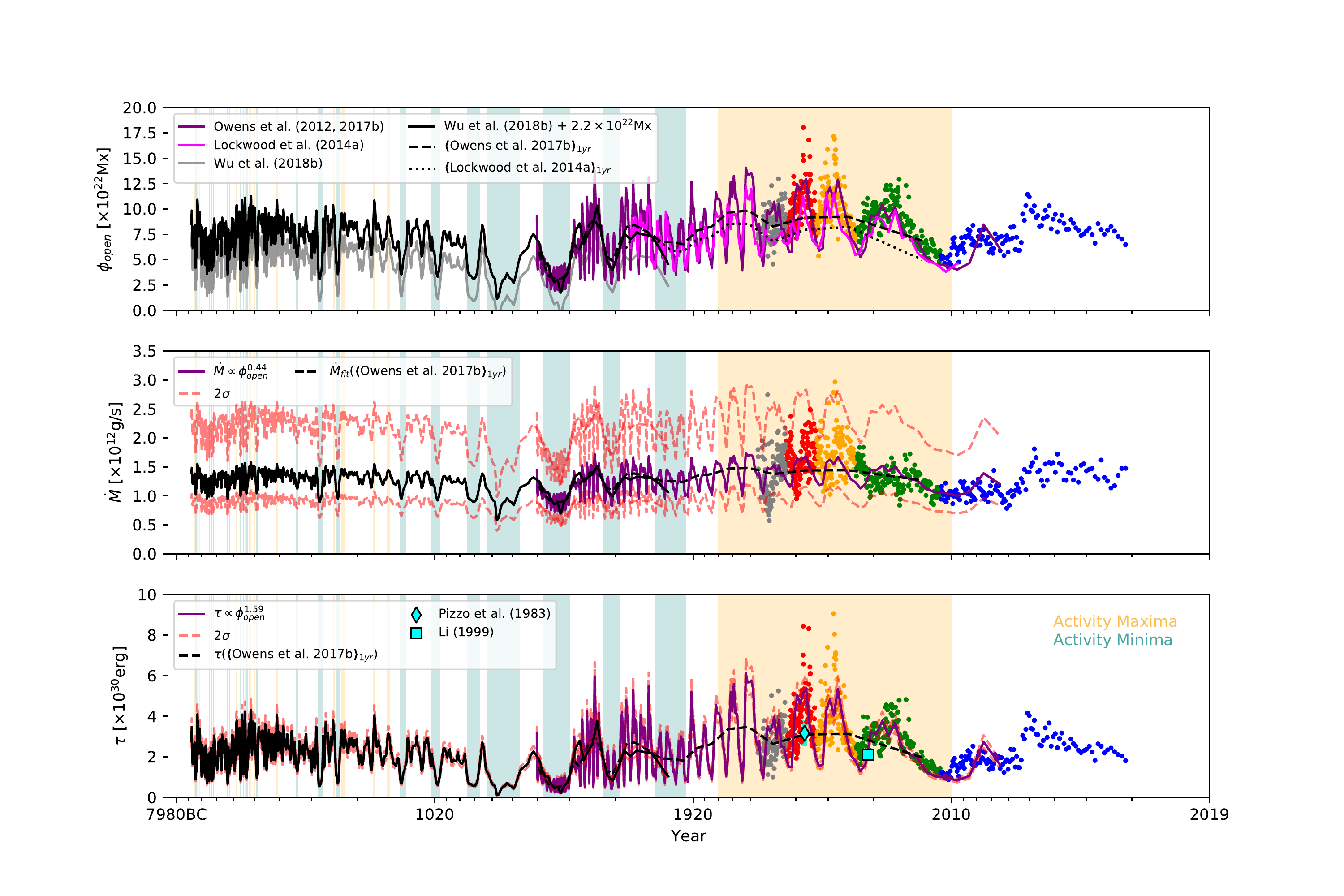}
  \caption{9000 years of solar open flux, $\phi_{open}$, mass loss rate, $\dot{M}$, and our predicted solar wind torque, $\tau$, versus inverse logarithmic look back time from 2019. The results derived from the OMNI dataset are plotted as they appeared in Figure \ref{OMNI_data}. The $\phi_{open}$ reconstructed by \cite{owens2017global} (group sunspot number) and \cite{lockwood2014reconstruction} (geomagnetic, aa-index) are plotted in the top panel with purple and magenta lines, respectively. We calibrate the long-time $\phi_{open}$ reconstruction from \cite{wusolar} (cosmogenic radionuclides), plotted in the top panel in grey, by first averaging the \cite{owens2017global} and \cite{lockwood2014reconstruction} reconstructions on the same decadal timescale, shown with dashed and dotted black lines respectively, then we shifted the \cite{wusolar} $\phi_{open}$ to match by adding a constant value. This reconstruction is shown with a solid black line, in good agreement with the smoothed values in the overlapping time period of $\sim1600-1900$. Using the $\phi_{open}$ from \cite{owens2017global} and \cite{wusolar}, the $\dot{M}$ predicted using equation (\ref{MLR_eq}) is plotted in the middle panel with solid purple and black lines respectively. The $\tau$ predicted by equation (\ref{open_torque2}), for each reconstruction is then plotted with solid purple and black lines in the bottom panel. For both predicted $\dot M$ and $\tau$, the 2$\sigma$ bound is indicated with dashed red lines. Maxima and minima in solar activity are shaded with color.}
  \label{mill}
\end{figure*}

For the centuries and millennia pre-dating the space age, estimates of the open magnetic flux have been produced using a number of different indirect methods.  {To compare them with indirect methods and over a wide range of timescales, we plot the spacecraft data from Figure~\ref{OMNI_data} also in Figure~\ref{mill}, which displays the solar wind parameters versus (inverse) logarithmic look-back time since 2019.}

\subsubsection{Centennial Variability}
{Geomagnetic disturbances, caused by the interaction of the solar wind and the Earth's magnetosphere, have been found to correlate well with solar activity, and thus the amount of open magnetic flux in the Heliosphere \citep{stamper1999solar, rouillard2007centennial, svalgaard2010heliospheric, lockwood2013reconstruction, lockwood2014centennial}. We plot the open magnetic flux reconstructed by \cite{lockwood2014reconstruction} using geomagnetic indices in the top panel of Figure \ref{mill} with a solid magenta line.
Additionally, the amount of open magnetic flux can be estimated from records of the observed sunspot number, which date back further than the records of the geomagnetic field \citep{solanki2002secular, krivova2007reconstruction, vieira2010evolution, owens2012cyclic}. We plot one such reconstruction from \cite{owens2012cyclic}, which is also used in \cite{owens2017global}, with a solid purple line in the top panel of Figure \ref{mill}.}

%


The two reconstructions (using geomagnetic and sunspot records), agree with each other and, during the space age, with the open magnetic flux from Section 3.1 as they were tuned by the authors to do so. These reconstructions reveal the behaviour of solar activity on a longer timescale than the 11 year sunspot cycle. It has been noted that during the last century the open magnetic flux has been at a sustained high with respect to the longer dataset \citep{lockwood2009rise}. Inspecting the past four centuries, there are also times when the open magnetic flux is shown to weaken for several magnetic cycles \citep{usoskin2015maunder}. We will examine the impact these different periods have on the solar wind torque in Section 4.3.

To examine the validity of {our approach},
we over-plot the reconstructed open magnetic flux (during the space-age) from \cite{owens2017global}, the mass loss rate {it predicts using}
equation (\ref{MLR_eq}), and the torque {it predicts} from equation (\ref{open_torque2}) in Figure \ref{OMNI_data} with solid black lines. Some temporal lag appears between the open magnetic flux and the observed mass loss rate, which is not captured in our prediction for the mass loss rate\footnote{We attempted to fit many different functions for $\dot M$, some of which considered a time-lag between $\dot M$ variations and the $\phi_{open}$. However, the additional complications did not statistically improve our $\dot M$ predictions. Therefore, we present a simple function of $\dot M(\phi_{open}$).}. Despite this, the 2$\sigma$ bounds of equation (\ref{MLR_eq}) roughly encompass the observed variation of the mass loss rate (as constructed). The predicted torque, from equation (\ref{open_torque2}), is found to be in good agreement with the torques calculated in Section 3.1. The 2$\sigma$ bound from the torque prediction, shown by red dashed lines, indicates a weak dependence of solar wind torque on the assumed mass loss rate. Therefore, provided the mass loss rate of the Sun has not changed significantly over each reconstructed timescale considered in this work, the open magnetic flux {alone} is capable of providing a good estimate of the solar wind torque.

\subsubsection{Millennial Variability}

To go back further the open magnetic flux can only be reconstructed using cosmogenic radionuclides. Cosmogenic radionuclides, such as \ce{^{14}C} and \ce{^{10}Be}, are produced as a byproduct of the interaction of galactic cosmic rays (GCRs) and the Earth's atmosphere.
This rate is modulated by the geomagnetic field, but also by features in the heliosphere, such as the interplanetary magnetic field and solar wind \citep{stuiver1961variations, stuiver1980changes}. Therefore, the concentration of cosmogenic radionuclides can be used as a proxy for solar variability \citep[see review][]{beer2012cosmogenic}.

\cite{wusolar} reconstructed the first solar modulation potential using multiple cosmogenic radionuclide records {(e.g., from tree rings for \ce{^{14}C}, and ice cores for \ce{^{10}Be})}, from which the solar open magnetic flux was calculated with a physics-based model \citep{wu2018solar}. We plot the open magnetic flux from \cite{wusolar} in the top panel of Figure \ref{mill} with a solid grey line. However, the values of the open magnetic flux appear too low where they overlap with the centennial reconstructions, and they sometimes contain negative values. This occurs as the generation of open magnetic flux is dependent on the reconstructed sunspot number, such that times when the modulation potential recovers zero sunspot number, they predict anomalously low values for the open magnetic flux. It is difficult to correctly account for this, so we will simply adjust this reconstruction to match the centennials reconstructions.
{To adjust the reconstructions of \cite{wusolar}, we create a comparison dataset by averaging} the open magnetic flux values from \cite{lockwood2014reconstruction} and \cite{owens2017global} on decadal timescales, to match the cadence recovered by the millennial reconstruction. These smoothed values are plotted with dotted and dashed lines respectively, in the top panel of Figure \ref{mill}. We then re-scale the reconstruction of \cite{wusolar} by adding a constant offset of $2.2\times10^{22}$Mx, shown with a solid black line, which brings the smoothed and millennial reconstructions into agreement. It is worth noting that {we have no physical justification for applying this linear shift to the reconstruction, which could introduce some (unknown) systematic error.}

Examining all the values of open magnetic flux collected in Figure \ref{mill}, the variability of the solar magnetic field appears to have a similar behaviour across a range of timescales. During the last several millennia, there appear to be times similar to the modern grand maxima, and the grand minima which are observed in the centennial reconstructions. We find no clear evidence for times of solar {open magnetic flux}
significantly greater than present in any of these records.

\subsection{Centuries and Millennia of Solar Wind Torque}
To evaluate the solar wind torque during the last four centuries we use the open magnetic flux from \cite{owens2017global}. In Figure \ref{mill}, we plot the mass loss rate using equation (\ref{MLR_eq}) and the resulting torque using equation (\ref{open_torque2}) with solid purple lines, and the 2$\sigma$ bounds with dashed red lines. The average solar wind torque during this ``centennial''-scale reconstruction is calculated to be $2.01\times10^{30}$erg. Similarly, in Figure \ref{mill} we plot the mass loss rate and torque using the ``millennial''-scale open flux reconstruction from \cite{wusolar} with solid black lines, along with the 2$\sigma$ bound in dashed red. We calculate the average torque for this dataset to be $2.16\times10^{30}$erg.
To better understand these results, we highlight historical maxima and minima of solar activity in Figure \ref{mill}, and evaluate the average torque for each of these time periods, where available. The dates for these are taken from the review of \cite{usoskin2017history} and are listed in Appendix Table \ref{HistoricalTorques}, along with their average torques.

Using the centennial reconstruction, the \textit{modern maximum} (which spans the majority of the 20th century), has a larger average torque of $3.14\times10^{30}$erg than considering the full centennial reconstruction. This is because the last four centuries also include multiple minima in solar activity, which host lower than average torques. Perhaps most notably the \textit{Maunder minimum} (which spans the years 1640 to 1720), which has an average torque of $0.67\times10^{30}$erg. Using the millennial reconstruction, we find the torque calculated during the \textit{Maunder minimum} is similar in strength to the many other named activity minima from the last 9000 years, such as the \textit{Sp\"{o}rer}, \textit{Wolf} and \textit{Oort} Minima. Reconstructions of solar activity appear to suggest the Sun spends around a sixth of its time in such a low torque state \citep[see][]{usoskin2007grand}, consistent with the \cite{wusolar} reconstruction. We find the solar wind torque during these activity minima have average values that span $0.62-1.73\times 10^{30}$erg, in contrast to the activity maxima which have much larger average values ranging from $2.44-3.87\times 10^{30}$erg.

Reconstructions of the solar open magnetic flux (or sunspot number), based on proxies of solar activity, allow for the detection of periodicities in the Sun's magnetic activity, on longer timescales than can be directly observed \citep{steinhilber20129, usoskin2016solar, wusolar}. Currently, there is little evidence for further variation, periodic or otherwise, in solar activity on longer timescales than the Hallstatt cycle which has a period of $\sim2400$ years \citep{sonett1991sun}. {Since}
the solar wind torque derived in this work is directly linked to solar activity, a similar conclusion can be made about the secular variation of the solar angular momentum loss rate.

\section{Discussion}
We have now calculated the solar wind torque on a variety of timescales. In this Section, we explore potential caveats to our results, and then compare our torques to those prescribed by models of the rotation period evolution of Sun-like stars.

\subsection{Reliability of Open Flux Proxies and Our Predicted Mass Loss Rates}
Indirect reconstructions of the solar open magnetic flux are by no means certain, and require careful examination and calibration. Geomagnetic indices (such as the aa-index) are often compiled from multiple ground-based monitoring stations, at differing latitudes in order to produce the most reliable value possible \citep[e.g.][]{clilverd2005reconstructing}. The interpretation of geomagnetic records as a proxy for open magnetic flux appears robust, at least for times where direct measurements are available for comparison \citep[see Figure 2 of][]{lockwood2004open}. Sunspot number records, from which our centennial torque is ultimately generated, often suffer from historical periods that are incomplete or uncertain due to a lack of reliable observers \citep{vaquero2011revisited, vaquero2014revised, munoz2018visualization}, or the modern interpretation of their recordings being under debate \citep[e.g.][]{usoskin2015maunder}. Models that recover the open magnetic flux based on sunspot number are shown to match concurrent geomagnetic and in-situ measurements where available \citep{solanki2002secular, vieira2010evolution, owens2012cyclic}. Our millennial torque is based on the changing concentration of cosmogenic radionuclides found in a range of terrestrial archives. This requires knowledge of the physical mechanisms which produce, transport and deposit each radioisotope  \citep[e.g.][]{reimer2009intcal09, heikkila2013atmospheric}. These processes typically smooth variability on decadal-timescales, such that the familiar 11 years sunspot cycle is not observed. Furthermore, linking these results to the open magnetic flux requires careful calibration \citep[e.g.][]{usoskin2003millennium, solanki2004unusual}.

The fact that the various proxies agree with each other where they overlap is because they were calibrated to do so. Typically the amplitude of variation in each reconstruction is a free parameter, but the waveform is fixed by the data. The implicit assumption made is that the relationship between each proxy and the open magnetic flux is the same in the past as it is now, though it is difficult to know whether these relationships may have changed during the timescale of each reconstruction. Despite the potential limitations of each reconstruction, we have taken each reconstruction at ``face value'' to characterise long-term variability, so our calculated torques carry all their associated uncertainties.

To reconstruct the mass loss rate of the Sun, we chose to fit equation (\ref{MLR_eq}) to the available data, and represented the apparent spread of values around this fit using a $2\sigma$ bound. The solar mass loss rate is not observed to vary substantially (extremes of $0.7-3.0\times10^{12}$g/s, see also \citealp{cohen2011independency}), and the torques calculated using equation (\ref{open_torque}) are weakly dependent on our choice of mass loss rate (when compared to the open magnetic flux). For example, to double the solar wind torque by only modifying the mass loss rate, would require the mass loss rate to increase by a factor of $\sim14$, therefore, unless the solar wind mass flux was very different in the past, uncertainties in the functional form of equation (\ref{Mdot_OSF}) do not significantly influence our results.

\subsection{Impacts of Magnetic Variability on Short Timescales}
Reconstructions of solar activity based on the concentrations of cosmogenic radionuclides incur smoothing effects from the transport and deposition timescales of each radionuclide. Therefore, such records struggle to recover short timescales variability, such as the 11 year sunspot cycle. Typically, this can be thought of as averaging the activity of the Sun over decadal timescales. Additionally, the centennial reconstruction is averaged on annual timescales and our in-situ measurements are averaged to 27 days. Due to the nonlinear dependence of equation (\ref{open_torque2}) on the open magnetic flux in the solar wind, short-term variability in the open magnetic flux, even around a fixed average value, will increase the long-term average torques. So, our millennial averaged torque using \cite{wusolar} is most likely slightly smaller than the true value.

The significance of this effect over the complete nine millennia can be probed in a few ways. {The standard deviation of the torque for each reconstruction about its average value is found to decrease as the averaging timescale grows. Consequently, each reconstruction is only sensitive to variability on timescales larger than the cadence of the dataset.} By comparing the average torques from the smoothed reconstructions of \cite{lockwood2014reconstruction} and \cite{owens2017global} to their original datasets, we find the original datasets have a larger torque by $\sim 4\%$ than their smoothed counterparts; a result of the non-linearity of the torque on open magnetic flux. For timescales shorter than 27 day, we have no measure of how variability affects our average values compared to the true value, {but observed variations on shorter timescales may be ever more dominated by spatial variations in the wind, rather than variations in the global, integrated wind properties.}

\subsection{Comparison to Rotation-Evolution Torques}

One motivation for the present work was the finding of \cite{finley2018effect}, that the solar wind torque is less than that predicted by a \cite{skumanich1972time} relation (a value of $6.2\times10^{30}$erg). One possible solution to this is that the torque varies on a longer timescale than the $\sim20$ years examined in that work. Here we rule out that variability on timescales of up to 9000 years can be the cause of this difference.  The average torque from the last nine millennia appears consistent with present-day torque calculations for the Sun \citep{pizzo1983determination, li1999magnetic, alvarado2016simulating,reville2017global,fionnagain2018solar,usmanov2018steady}.  {In order to reconcile the solar wind torque with that predicted by the Skumanich relation, the average open magnetic flux, for example, would need to be $\sim 14\times10^{22}$ Mx, which is well above most measurements shown in the top panel of Figure~\ref{mill}.}

However, we cannot rule out {that the torque varies on longer timescales.  Any cyclical variations in the torque on timescales shorter than $\sim 10^{7}-10^{8}$ years would not noticeably change the observed spin distributions of stars with ages $\ga 1$ Gyr.  Thus, the solar torque could still be reconciled with the Skumanich torque, if it varies on much longer timescales than probed here, and if the sun is currently in a ``low torque state.''  Alternatively, if the estimates of the present-day solar wind torque are correct, they may be consistend with the suggestion of \citet{van2016weakened} that sun-like stars transition to a state of permanently weakened torque at approximately the solar age.  If that is the case, our results mean that this transition either occured more than $\sim 10^4$ years ago for the sun, or that any continuing transition is so gradual as not to be measureable on that timescale.}



{If the solar wind torque does indeed vary significantly on longer timescales than probed here, it suggests that the present-day wind torques of other stars should scatter (by at least a factor of $\sim$3) around the torque predicted by rotation-evolution models.  Recently, \citet{finley2019effect} estimated the torques of 4 stars that had surface magnetic field measurements and some information about their mass loss rates (see also See et al., submitted).  In all cases, the estimated torques were a factor of several times smaller than inferred from rotation-evolution models.  They only studied 4 stars, and the systematic uncertainties are large, but this is evidence against significant long-term cyclical variability causing the discrepancy.}


If {long-term} variability in the angular momentum loss rate of Sun-like stars does not resolve this discrepancy, then {it could indicate systematic errors in}
the wind models, or the observed wind parameters{, although the origins of such errors are unclear}.
{On the shortest timescales,} there also exist a range of transient phenomena in the corona \citep{cane2003interplanetary, rod2016effect, sanchez2017temporal}, along with short-timescale variations in the solar wind \citep{king2005solar, thatcher2011statistical}, which are not incorporated into steady state solutions of the wind. The impact these have on our semi-analytic formulae for the torque (i.e., equation (\ref{open_torque})) are poorly constrained \citep{aarnio2012mass}.

\section{Conclusion}
In this paper we have investigated the angular momentum loss rate of the Sun on a longer timescale than previously attempted. To do this, we use the semi-analytic braking law of \citetalias{finley2018dipquadoct} to calculate the torque on the Sun due to the solar wind. We first expand the calculation of \cite{finley2018effect} throughout the entire space-age by using in-situ spacecraft measurements, taken from the OMNI dataset. We then utilise reconstructions of the solar open magnetic flux, based on geomagnetic indices \citep{lockwood2014reconstruction}, sunspot number records \citep{owens2012cyclic}, and concentrations of cosmogenic radionuclides \citep{wusolar}, to estimate the braking torque over the last four centuries, and then the last nine millennia.

The Sun undergoes significant variation in its magnetic activity on centennial and millennial timescales, which include times of grand maxima and minima of activity. The average torque during grand maxima ranges from $2.4-3.9\times10^{30}$erg, with peaks of $\sim5\times 10^{30}$erg. To contrast this, grand minima (such as the \textit{Maunder}, \textit{Sp\"{o}rer}, \textit{Wolf} and \textit{Oort} minimum) produce some of the lowest values from $0.6-1.7\times10^{30}$erg. Overall, we find the average angular momentum loss rate of the Sun, during the last nine millennia to be $2.2\times10^{30}$erg, which is equal to the average value during the last two decades.

The values calculated in this work remain contrary to those required by current rotation-evolution models of Sun-like stars. Such models predict a braking torque of $6.2\times10^{30}$erg \citep{matt2015mass, finley2018effect}, which we do not recover by using data spanning from present to 6755BC, roughly 9000 years. This discrepancy could be due to the simplicity of the current MHD wind models, or to much longer timescale variation in the solar torque, or to uncertainties in measuring solar wind parameters (and inferring them in the past), or to significant deviations in the spin-down torque of low-mass stars from the \cite{skumanich1972time} relation around the age of the Sun. Further exploration of this discrepancy is required, and with Parker Solar Probe making in-situ measurements of the solar wind closer to the Sun than previously attempted \citep{fox2016solar}, a direct measurement of the angular momentum loss rate would {help}
to validate, or discredit, our calculations.

\acknowledgments
We thank the many instrument teams whose data contributed to the OMNI dataset, and the NASA/GSFC's Space Physics Data Facility's OMNIWeb service for providing this data.
AJF, SD and SPM acknowledge funding from the European Research Council (ERC) under the European Union’s Horizon 2020 research and innovation programme (grant agreement No 682393 AWESoMeStars).
MO is funded by Science and Technology Facilities Council (STFC) grant numbers ST/M000885/1 and ST/R000921/1
Figures in this work are produced using the python package matplotlib \citep{hunter2007matplotlib}.

\appendix
\section{Grand Maxima and Minima Solar Wind Torques}
For the solar angular momentum loss rate generated using equation (\ref{open_torque2}) and the open magnetic flux reconstructions of \cite{owens2012cyclic} and \cite{wusolar}, centennial and millennial-scale reconstructions respectively, we {list in Table \ref{HistoricalTorques}}
the average values during historical grand maxima and minima in solar activity. The dates for which are taken from the review of \cite{usoskin2017history}.

\begin{table}
\caption{Average Solar Wind Torques for Historical Activity Maxima and Minima}
\label{HistoricalTorques}
\center
\setlength{\tabcolsep}{3pt}
  \begin{tabular}{cccccc}
      \hline\hline
&	Period	&	\multicolumn{2}{c}{Duration}			&	\multicolumn{2}{c}{$\langle\tau\rangle$}			\\
&	Name	&	\multicolumn{2}{c}{Year (-BC/AC)}			&	\multicolumn{2}{c}{($\times 10^{30}$erg)}			\\
\cmidrule(lr){3-4}	\cmidrule(lr){5-6}
&		&	Start	&	End	&	Centennial	&	Millennial	\\	\hline
&	Modern 	&	1940	&	2010	&	3.14	&	-	\\
&	-	&	480	&	530	&	-	&	2.91	\\
&	-	&	290	&	330	&	-	&	3.67	\\
&	-	&	-280	&	-210	&	-	&	2.93	\\
&	-	&	-460	&	-410	&	-	&	3.26	\\
&	-	&	-2090	&	-2040	&	-	&	3.68	\\
&	-	&	-2970	&	-2940	&	-	&	3.47	\\
&	-	&	-3220	&	-3120	&	-	&	3.80	\\
&	-	&	-3430	&	-3380	&	-	&	3.72	\\
\rot{\rlap{~Activity Maxima}}
&	-	&	-3885	&	-3835	&	-	&	3.23	\\
&	-	&	-6140	&	-6100	&	-	&	2.95	\\
&	-	&	-6300	&	-6260	&	-	&	3.87	\\
&	-	&	-6550	&	-6480	&	-	&	2.44	\\
&	-	&	-6730	&	-6690	&	-	&	3.09	\\	\hline
&	Glassberg 	&	1880	&	1914	&	2.22	&	-	\\
&	Dalton 	&	1797	&	1828	&	1.22	&	1.11	\\
&	Maunder 	&	1640	&	1720	&	0.67	&	0.67	\\
&	Sp\"{o}rer 	&	1390	&	1550	&	-	&	0.62	\\
&	Wolf 	&	1270	&	1350	&	-	&	0.74	\\
&	Oort  	&	990	&	1070	&	-	&	1.20	\\
&	-	&	650	&	730	&	-	&	1.12	\\
&	-	&	-400	&	-320	&	-	&	1.04	\\
&	-	&	-810	&	-690	&	-	&	0.89	\\
&	-	&	-1400	&	-1350	&	-	&	1.46	\\
&	-	&	-2470	&	-2430	&	-	&	1.73	\\
&	-	&	-2900	&	-2810	&	-	&	1.30	\\
&	-	&	-3370	&	-3280	&	-	&	1.21	\\
\rot{\rlap{~Activity Minima}}
&	-	&	-3520	&	-3470	&	-	&	1.39	\\
&	-	&	-3645	&	-3595	&	-	&	1.18	\\
&	-	&	-4235	&	-4205	&	-	&	1.42	\\
&	-	&	-4340	&	-4290	&	-	&	1.05	\\
&	-	&	-5220	&	-5170	&	-	&	1.25	\\
&	-	&	-5325	&	-5275	&	-	&	1.20	\\
&	-	&	-5480	&	-5440	&	-	&	1.03	\\
&	-	&	-5630	&	-5590	&	-	&	0.93	\\
&	-	&	-6450	&	-6320	&	-	&	1.33	\\
\hline
  \end{tabular}
\end{table}




\bibliographystyle{yahapj}
\bibliography{Adam}



\end{document}